\documentclass[aps, pra, amsfonts, amsmath, amssymb, superscriptaddress, floatfix]{revtex4}

\usepackage{graphicx}
\usepackage{dcolumn}
\usepackage{bm}

\begin{document}
\title{Atom and photon measurement in cooperative scattering by cold atoms}

\author{Tom Bienaim\'e}
\affiliation{Universit\'e de Nice Sophia Antipolis, CNRS, Institut Non-Lin\'eaire de Nice, UMR 7335, F-06560 Valbonne, France}

\author{Marco Petruzzo}
\affiliation{Dipartimento di Fisica, Universit\`a Degli Studi di Milano, Via Celoria 16, I-20133 Milano, Italy}

\author{Daniele Bigerni}
\affiliation{Dipartimento di Fisica, Universit\`a Degli Studi di Milano, Via Celoria 16, I-20133 Milano, Italy}

\author{Nicola Piovella}
\affiliation{Dipartimento di Fisica, Universit\`a Degli Studi di Milano, Via Celoria 16, I-20133 Milano, Italy}

\author{Robin Kaiser} \email{robin.kaiser@inln.cnrs.fr}
\affiliation{Universit\'e de Nice Sophia Antipolis, CNRS, Institut Non-Lin\'eaire de Nice, UMR 7335, F-06560 Valbonne, France}

\begin{abstract}

In this paper, we study cooperative scattering of low intensity
light by a cloud of N two-level systems. We include the incident
laser field driving these two-level systems and compute the
radiation pressure force on the center of mass of the cloud. This
signature is of particular interest for experiments with laser
cooled atoms. Including the complex coupling between dipoles in a
scalar model for dilute clouds of two-level systems, we obtain
expression for cooperative scattering forces taking into account
the collective Lamb shift. We also derive the expression of the
radiation pressure force on a large cloud of two-level systems
from an heuristic approach and show that at lowest driving
intensities this force is identical for a product and an entangled
state.

\end{abstract}

\maketitle

\section{Introduction}\label{Intro}

Cooperative scattering by an assembly of resonant systems has been studied in detail for many years and is based on the seminal work by R.
Dicke in 1954 \cite{Dicke54}. Related superradiance effects and collective level shifts have been studied in the context of atomic physics
in the 70s \cite{Lehmberg68, Friedberg,Gross82}. In the last decade, this topic has seen renewed interest
\cite{Eberly06, Scully06, Friedberg07, Svidzinsky08, Scully09, Scully09LS, Svi10, Friedberg10,Friedberg10b,Prasad10} with  novel experiments
in nuclear physics \cite{Rohlsberger10} and in laser cooled clouds of atoms \cite{Bienaime10, Bender10, Courteille10, Kaiser09, Bux10}, applications in
quantum information \cite{Greentree06} and quantum phase transitions \cite{Osterloh02, Akkermanns08}. As we are mainly concerned with applications
on laser cooled atomic samples, we focus in this paper on specific parameters and observables which are of interest in such experiments.
We therefore derive expressions of the radiation pressure force acting on
the center of mass of the atomic cloud, as well as the scattered electric
field. We go beyond past approximations including the complex kernel for
the coupling terms between N atoms \cite{Friedberg, Svidzinsky08}, described by two-level systems in
a scalar approach. Neglecting the complete vectorial nature of the dipole
dipole coupling seems a priori more justified in a dilute sample of atoms,
where near field corrections are small \cite{Kaiser09}. Furthermore, we obtain the
force and the radiation field as quantum operators, which may be useful
for studying fluctuations and diffusion effects in radiation forces and
scattered emission. Also, the imaginary part of the complex kernel,
describing the collective Lamb shift, is evaluated for a gaussian density
profile.

This paper is organized as follows: in section \ref{Hamiltonian}, we specify the Hamiltonian used and discuss our approximations.
In section \ref{Observables}, we introduce the observables relevant for experiments with cold atoms, namely the radiation pressure forces on the
center of mass of the atomic cloud and the scattered light intensity. The evaluation of these observables is done for specific atomic states in
section \ref{AtomicState}.
We derive the result for this cooperative radiation pressure force from a more heuristic approach in section \ref{HeuristicApproach}.
In section \ref{ProductState} we discuss the relevance of the Timed Dicke State compared to a product state for this cooperative pressure force
in the low intensity limit before concluding in section \ref{Conclusion}.

\section{Hamiltonian and operator equations}\label{Hamiltonian}
\begin{figure}[t]
        \centerline{{\includegraphics[height=6cm]{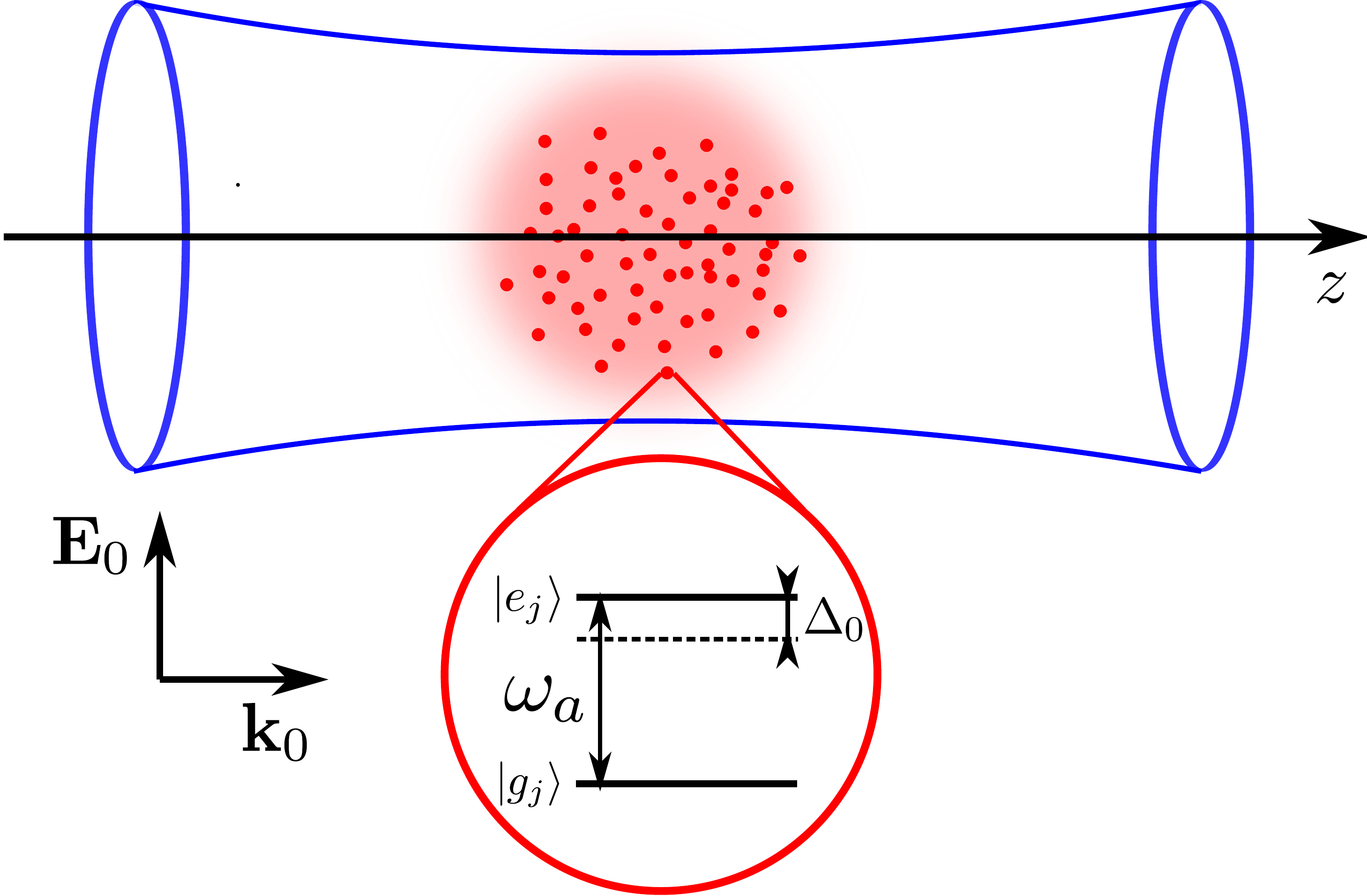}}}
        \caption{(color online) Experimental configuration considered : a cloud of two-level atoms is driven by an incident laser
        detuned by $\Delta_0$ from the atomic resonance $\omega_a$, with wavevector $\mathbf{k}_0$.}
        \label{Fig1}
    \end{figure}

Our system  consists of a gas of $N$ two-level atoms (with random
positions $\mathbf{r}_j$, lower and upper states $|g_j\rangle$ and
$|e_j\rangle$ with $j=1,\dots,N$, transition frequency $\omega_a$
with linewidth $\Gamma=d^2\omega_a^3/2\pi\hbar\epsilon_0 c^3$, where
$d$ is the electric dipole matrix element), driven by a uniform
resonant radiation beam  with wave vector
$\mathbf{k}_0=k_0\mathbf{\hat e}_z$, frequency
$\omega_0=\omega_a+\Delta_0$ and electric field $E_0$ (see fig. \ref{Fig1}).

The atom-field interaction Hamiltonian in the rotating-wave approximation (RWA) is
\begin{equation}\label{H}
\hat H=\hat H_0+\hat H_1
\end{equation}
where
\begin{eqnarray}
    \hat{H}_0 &=&\hbar\sum_{j=1}^N\left\{
    -\frac{\Delta_0}{2}\hat\sigma_{3j}+\frac{\Omega_0}{2}
    \left(
    \hat\sigma_je^{-i\mathbf{k}_0\cdot\mathbf{r}_j}
    +{\hat\sigma_j}^{\dagger}e^{i\mathbf{k}_0\cdot\mathbf{r}_j}\right)\right\}\label{H0}\nonumber\\
    \hat{H}_1 &=&\hbar\sum_{j=1}^N\sum_{\mathbf{k}}g_k\left[\hat{a}_{\mathbf{k}}^\dagger\hat\sigma_j
    e^{i(\omega_k-\omega_0)t-i\mathbf{k}\cdot\mathbf{r}_j}
        +{\hat\sigma_j}^{\dagger}\hat{a}_{\mathbf{k}}
    e^{-i(\omega_k-\omega_0)t+i\mathbf{k}\cdot\mathbf{r}_j}\right]\label{H1}.
\end{eqnarray}
Here $\Omega_0=dE_0/\hbar$ is the pump Rabi frequency,
$\hat{a}_{\mathbf{k}}$ is the photon annihilation operator with
wavenumber $\mathbf{k}$ and frequency $\omega_k=ck$, $g_k
=d\sqrt{\omega_k/(2\hbar\epsilon_0V_{ph})}$, $V_{ph}$ the photon
volume, $\hat\sigma_j=\exp(i\Delta_0 t)|g_j\rangle\langle e_j|$
and $\hat\sigma_{3j}=|e_j\rangle\langle e_j|-|g_j\rangle\langle
g_j|$. Instead of solving the Schr\"{o}dinger equation introducing
some ansatz for the system state $|\Psi(t)\rangle$ \cite{Courteille10}, we
write the motion equations of the atomic and field operators,
\begin{eqnarray}
  \frac{d\hat\sigma_{j}}{dt} &=& \frac{1}{i\hbar}[\hat\sigma_j,\hat H]=
  i\Delta_0\hat\sigma_j+\frac{i\Omega_0}{2}\hat\sigma_{3j}e^{i\mathbf{k}_0\cdot\mathbf{r}_j} +i\sum_{\mathbf{k}}g_k
  \hat\sigma_{3j}\hat a_{\mathbf{k}}e^{-i(\omega_k-\omega_0)t+i\mathbf{k}\cdot \mathbf{r}_j}\label{s1bis}\\
  \frac{d\hat\sigma_{3j}}{dt} &=& \frac{1}{i\hbar}[\hat\sigma_{3j},\hat H]=
   i\Omega_0 \hat\sigma_{j}e^{-i\mathbf{k}_0\cdot\mathbf{r}_j}+2i\sum_{\mathbf{k}}g_k
   \hat a_{\mathbf{k}}^\dagger\sigma_{j} e^{i(\omega_k-\omega_0)t-i\mathbf{k}\cdot
  \mathbf{r}_j}+ \textrm{h.c.}\label{s3bis}\\
  \frac{d\hat a_{\mathbf{k}}}{dt} &=& \frac{1}{i\hbar}[\hat a_{\mathbf{k}},\hat H]=
  -ig_ke^{i(\omega_k-\omega_0)t}\sum_{m=1}^N \hat\sigma_{m}e^{-i\mathbf{k}\cdot
  \mathbf{r}_m}\label{akbis}.
\end{eqnarray}
We consider the atoms initially in their ground state and we
assume weak excitation ($\Omega_0\ll\Gamma$), so that we
approximate $\hat\sigma_{3j}(t)\approx -\hat I_j$, where $\hat I_j$ is
the identity operator for the $j$th atom. This approximation amounts to neglect
saturation and multi-excitation, i.e. all the processes generating
more than one photon at the same time (\textit{linear regime}).
Integrating Eq.(\ref{akbis}) and substituting it into
Eq.(\ref{s1bis}), neglecting $a_k(0)$ (since the initial field
state is vacuum) we obtain
\begin{eqnarray}
  \frac{d\hat\sigma_{j}}{dt} &=&
  i\Delta_0\hat\sigma_j-\frac{i\Omega_0}{2} \hat I_j e^{i\mathbf{k}_0\cdot\mathbf{r}_j}-
  \sum_{\mathbf{k}}g_k^2\sum_{m=1}^N
  e^{i\mathbf{k}\cdot(\mathbf{r}_j-\mathbf{r}_m)}
  \int_0^t dt'
  \hat\sigma_m(t-t')\,e^{-i(\omega_k-\omega_0)t'}\label{s1ter}.
\end{eqnarray}
The last term in Eq.(\ref{s1ter}) describes the effect of the
spontaneously emitted photons on the atoms, and it is well known
in the quantum electrodynamic literature \cite{Scully:QO,Agarwal}.
In the Markov approximation (i.e.
when the photon transit time through the atomic sample is much shorter
than the excitation decay time \cite{velocity}), we assume under
the integral $\hat\sigma_m(t-t')\approx \hat\sigma_m(t)$.
The time integral then yields a real part (with a term  $\delta (k-k_0)$ )
and an imaginary part (corresponding to the principal part of the integral).
Taking into account these two terms is at the origin of the exponential kernel whereas the real part alone would lead to a sin
kernel in Eq.(\ref{kernel}) below.
We then transform the sum over the modes $\mathbf{k}$ into an integral,
$\sum_{\mathbf{k}}\rightarrow (V_{ph}/8\pi^3)\int d\mathbf{k}$.
The real and imaginary parts of the double integral over $t$
and $\mathbf{k}$ yield the cooperative decay and frequency shift
(collective Lamb shift), respectively. The proper expression of
the cooperative frequency shift has been obtained  adding to the
Hamiltonian (\ref{H1}) the not-RWA contributions associated to
virtual photons exchanged between different atoms. It results the
following relation \cite{Svi10}:
\begin{equation}\label{rel:kern}
\sum_{\mathbf{k}}g_k^2
  e^{i\mathbf{k}\cdot
   \mathbf{R}}
  \int_0^\infty dt'e^{-ic(k-k_0)t'}\longrightarrow \frac{\Gamma}{2ik_0
  |\mathbf{R}|}e^{ik_0 |\mathbf{R}|}
\end{equation}
where $\Gamma=V_{ph}g_{k_0}^2 k_0^2/(\pi c)$.
Using Eq.(\ref{rel:kern}) in Eq.(\ref{s1ter}) we obtain
\cite{Svi10},
\begin{eqnarray}
   \frac{d\hat\sigma_{j}(t)}{dt} &=&
  i\Delta_0\hat\sigma_j(t)-\frac{i\Omega_0}{2}\hat I_j e^{i\mathbf{k}_0\cdot\mathbf{r}_j}-\frac{\Gamma}{2}
  \sum_{m=1}^N\gamma_{jm}
  \hat\sigma_m(t)\label{s1:Mark}.
\end{eqnarray}
where
\begin{equation}
\gamma_{jm}=\frac{-i\cos(k_0r_{jm})+\sin(k_0r_{jm})}{k_0r_{jm}}
=\frac{e^{ik_0r_{jm}}}{ik_0 r_{jm}}. \label{kernel}
\end{equation}
and $r_{jm}=|\mathbf{r}_j-\mathbf{r}_m|$. Eqs.(\ref{s1:Mark})
describe the time evolution of the atomic operators for $N$ weakly excited atoms scattering
radiation. The real part of $\gamma_{jm}$ describes the
spontaneous emission decay and the imaginary part of $\gamma_{jm}$
describes the energy shift due to resonant dipole-dipole
interactions.
A slightly different approach can be used to derive this result as shown in appendix \ref{AppendixA}.
Note that even though this result will yield a density dependent collective shift of the resonance,
we use a scalar model for the field, neglecting thus any polarization and near field dependence \cite{Scully09LS,Friedberg10}.
Detailed calculations for small and large samples of various geometries however show that near field and far field contributions as well
as resonant and antiresonant terms need to be taken properly into account for quantitative predictions \cite{Friedberg, Friedberg10, Friedberg10b},
and the present model thus needs to be considered with care illustrating only a part of the dipole-dipole coupling for real systems.

Eq.(\ref{s1:Mark}) can also cast in the form
\begin{eqnarray}
   \frac{d\hat\sigma_{j}}{dt} &=&\frac{1}{i\hbar}[\hat \sigma_j,\hat
   H_0'+\hat H_{eff}]\label{eqHeff},
\end{eqnarray}
where
\begin{eqnarray}
    {\hat{H}_0}' &=&\hbar\sum_{j=1}^N\left\{
    -\Delta_0\hat\sigma^\dagger_{j}\hat\sigma_{j}+\frac{\Omega_0}{2}
    \left(
    \hat\sigma_je^{-i\mathbf{k}_0\cdot\mathbf{r}_j}
    +{\hat\sigma_j}^{\dagger}e^{i\mathbf{k}_0\cdot\mathbf{r}_j}\right)\right\}\label{H0}\nonumber\\
    \hat H_{eff}&=&\frac{\hbar\Gamma}{2}\sum_{j,m}\frac{e^{ik_0r_{jm}}}{ik_0r_{jm}}
    \hat{\sigma}_j^\dagger\hat\sigma_m.\label{Heff}
\end{eqnarray}
and the commutation rules in the \textit{linear regime} are
$[\hat\sigma_j,\hat{\sigma}_m^\dagger]=\delta_{jm}$.

\section{Observables}\label{Observables}
 Among the different
observables of the system, scattered light and radiation pressure
force contain important signatures of cooperative scattering.
Concerning scattered radiation, the positive-frequency part of the
electric field is defined as
\begin{equation}\label{E}
    \hat E(\mathbf{r},t)=i\sum_{\mathbf{k}}{\cal
    E}_{k}\hat a_\mathbf{k}(t) e^{-i\omega_k t+i\mathbf{k}\cdot\mathbf{r}}
\end{equation}
where
${\cal E}_{k}=\sqrt{\hbar\omega_k/2\epsilon_0 V_{ph}}$
is the
single-photon electric field. By integrating Eq.(\ref{akbis}) and
inserting it in Eq.(\ref{E}) we obtain
\begin{equation}\label{E2}
    \hat E(\mathbf{r},t)=\sum_{\mathbf{k}}{\cal
    E}_{k}g_k\sum_{m=1}^N e^{i\mathbf{k}\cdot(\mathbf{r}-\mathbf{r}_m)-i\omega_0 t}
    \int_0^t dt'e^{-i(\omega_k-\omega_0)t'}\hat\sigma_m(t-t')
\end{equation}
Using Eq.(\ref{rel:kern}), the Markov approximation leads to
\begin{equation}\label{E3}
    \hat E(\mathbf{r},t)\approx -i\frac{d k_0^2}{4\pi\epsilon_0}\sum_{j=1}^N
    \frac{e^{-i\omega_0(t-|\mathbf{r}-\mathbf{r}_j|/c)}}{|\mathbf{r}-\mathbf{r}_j|}
    \hat\sigma_j(t)
\end{equation}
which has a transparent interpretation as the sum of wavelets
scattered by $N$ dipoles of position $\mathbf{r}_j$ and detected
at distance $\mathbf{r}$ and time $t$. In the far field
limit, $|\mathbf{r}-\mathbf{r}_j|\approx r-(\mathbf{r}\cdot
\mathbf{r}_j)/r$ and
\begin{equation}\label{E4}
    \hat E(\mathbf{r},t)\approx -i\frac{d k_0^2}{4\pi\epsilon_0 r}e^{-i\omega_0(t-r/c)}\sum_{j=1}^N
    e^{-i\mathbf{k_{s}}\cdot \mathbf{r}_j}
    \hat\sigma_j(t)
\end{equation}
where $\mathbf{k}_s=k_0(\mathbf{r}/r)$.

The radiation pressure force acting on the $j$th-atom has been
calculated from Eq.(\ref{H}) as
$\hat{\mathbf{F}}_j=-\nabla_{\mathbf{r}_j}\hat
H=\hat{\mathbf{F}}_{aj}+\hat{\mathbf{F}}_{ej}$  where \cite{Courteille10}
\begin{eqnarray}
    \hat{\mathbf{F}}_{aj}&=& i\hbar \mathbf{k}_0\frac{\Omega_0}{2}
    \left\{e^{-i\mathbf{k}_0\cdot \mathbf{}r_j}\hat\sigma_{j}-
    \textrm{h.c.}\right\}\label{Force-abs}\\
   \hat{\mathbf{F}}_{ej}&=& i\hbar\sum_{\mathbf{k}} \mathbf{k}g_{k}
    \left\{\hat a_{\mathbf{k}}^\dagger \hat\sigma_{j}
    e^{i(\omega_k-\omega_0)t-i\mathbf{k}\cdot\mathbf{r}_j}-
    \hat\sigma_{j}^\dagger \hat a_{\mathbf{k}}
    e^{-i(\omega_k-\omega_0)t+i\mathbf{k}\cdot\mathbf{r}_j}\right\}
    \label{Force-emi}
\end{eqnarray}
where $\hat{\mathbf{F}}_{aj}$ and $\hat{\mathbf{F}}_{ej}$ result
from the recoil received upon absorption of a photon from the pump
and from the emission of a photon into any direction $\mathbf{k}$,
respectively. Eliminating the field using Eq.(\ref{akbis}),
Eq.(\ref{Force-emi}) becomes
\begin{eqnarray}
   \hat{\mathbf{F}}_{ej}(t)= -\hbar\sum_{\mathbf{k}}
   \mathbf{k}g_{k}^2
    &\,&\left\{\sum_{m=1}^N
    e^{-i\mathbf{k}\cdot(\mathbf{r}_j-\mathbf{r}_m)}\int_0^t dt'
    e^{i(\omega_k-\omega_0)t'}\hat\sigma_m^\dagger(t-t')\,\hat\sigma_{j}(t) \right.\nonumber\\
    &\,&+\left.
    \hat\sigma_{j}^\dagger(t) \sum_{m=1}^N
    e^{i\mathbf{k}\cdot(\mathbf{r}_j-\mathbf{r}_m)}\int_0^t dt'
    e^{-i(\omega_k-\omega_0)t'}\hat\sigma_m(t-t')
    \right\}.
    \label{Force-emi:2}
\end{eqnarray}
Assuming the Markov approximation, $\hat\sigma_m(t-t')\approx
\hat\sigma_m(t)$, then Eq.(\ref{Force-emi:2}) becomes
\begin{eqnarray}
   \hat{\mathbf{F}}_{ej}(t)&=& -\hbar\sum_{m=1}^N\sum_{\mathbf{k}}
   \mathbf{k}g_{k}^2
    \left\{
    \hat\sigma_m^\dagger(t) \hat\sigma_{j}(t)
    e^{-i\mathbf{k}\cdot\mathbf{r}_{jm}}\int_0^t dt'
    e^{i(\omega_k-\omega_0)t'}
    +
    \hat\sigma_{j}^\dagger(t)\hat\sigma_m(t)
    e^{i\mathbf{k}\cdot\mathbf{r}_{jm}}\int_0^t dt'
    e^{-i(\omega_k-\omega_0)t'}
    \right\}\label{Fe:1}
\end{eqnarray}
where $\mathbf{r}_{jm}=\mathbf{r}_j-\mathbf{r}_m$. The force
(\ref{Fe:1}) acting on the $j$th atom has a single-atom
contribution $\hat{\mathbf{F}}_{ej}^{(\textrm{self})}$ (term $m=j$
in the sum) accounting for its own photon emission recoil, and a
contribution $\hat{\mathbf{F}}_{ej}^{(\textrm{int})}$ (terms
$m\neq j$) accounting for coupling between the $j$th
atom and all the other atoms. Note that this dipole-dipole interaction can occur via a coupling to common vacuum modes of radiation.
The interference terms in the total scattered field can leave a fingerprint on the forces acting on the atoms inside the cloud. The
first contribution yields
\begin{eqnarray}
   \hat{\mathbf{F}}_{ej}^{(\textrm{self})}&\approx & -\hbar\Gamma\sum_{|\mathbf{k}|=k_0}
   \mathbf{k}\,
    \hat\sigma_{j}^\dagger\hat\sigma_j,\label{self}
\end{eqnarray}
where the sum is over all the randomly oriented modes
$\mathbf{k}=k_0\hat{\mathbf{k}}$ and we have
omitted the self-energy shift (Lamb shift) coming
from the principal part term of the time integral
in Eq(\ref{Fe:1}). Noting that
for $m\neq j$ we have $i\mathbf{k}\exp(i\mathbf{k}\cdot
\mathbf{r}_{jm})=\nabla_{\mathbf{r}_{j}}\exp(i\mathbf{k}\cdot
\mathbf{r}_{jm})$, the second contribution to Eq.(\ref{Fe:1}) can
be written as
\begin{eqnarray}
    \hat{\mathbf{F}}_{ej}^{(\textrm{int})}(t) &=& -i\hbar\nabla_{\mathbf{r}_{j}}\sum_{m\neq j} \sum_{\mathbf{k}}
   g_{k}^2
    \left\{\hat\sigma_{j}(t)\hat\sigma_m^\dagger(t)
    e^{-i\mathbf{k}\cdot\mathbf{r}_{jm}}\int_0^t dt'
    e^{i(\omega_k-\omega_0)t'}-
    \textrm{h.c.}
    \right\}.
    \label{Force-emi:3}
\end{eqnarray}
 Using Eq.(\ref{rel:kern}) in Eq.(\ref{Force-emi:3}),
 Eq.(\ref{Fe:1}) becomes
\begin{eqnarray}
   \hat{\mathbf{F}}_{ej}(t) &=&
   \hat{\mathbf{F}}_{ej}^{(\textrm{self})}(t) -\nabla_{\mathbf{r}_{j}}\sum_{m\neq j}\hat V_{jm}(t).
    \label{Force-emi:4}
\end{eqnarray}
where
\begin{equation}\label{Vjm}
    \hat V_{jm}(t)= -\frac{\hbar\Gamma}{2}
    \left\{\frac{\hat\sigma^\dagger_{j}(t)\hat\sigma_m(t)e^{-ik_0r_{jm}}+
    \hat\sigma_j(t)\hat\sigma_{m}^\dagger(t)e^{ik_0r_{jm}}}{k_0
    r_{jm}}\right\}
\end{equation}
is the effective interaction energy between jth and mth atoms. Since
$\nabla_{\mathbf{r}}[\exp(ik_0 r)/r]=\mathbf{r}(ik_0 r-1)\exp(ik_0
r)/r^3$, Eq.(\ref{Force-emi:4}) becomes
\begin{eqnarray}
   \hat{\mathbf{F}}_{ej} &=&\hat{\mathbf{F}}_{ej}^{(\textrm{self})}
    -\frac{\hbar k_0\Gamma}{2}\sum_{m=1}^N
   \frac{\hat{\mathbf{n}}_{jm}}{(k_0 r_{jm})^2}
    \left\{\hat\sigma^\dagger_{j}\hat\sigma_m(1+ik_0r_{jm})e^{-ik_0r_{jm}}+h.c.\right\},
    \label{Force-emi:5}
\end{eqnarray}
where $\hat{\mathbf{n}}_{jm}=\mathbf{r}_{jm}/r_{jm}$. The emission
force acting on the $j$th atom has two contributions: a) a
self-force due to its own photon emission; b) a force due to the
dipole-dipole interactions with all the other atoms.
This second force has a term
decreasing as $1/r_{jm}$ and one decreasing as
$1/{r_{jm}}^2$.
\section{Atomic state}\label{AtomicState}
The linear approximation assumed in the
equations of the atomic operators $\hat\sigma_j$,
Eq.(\ref{s1:Mark}), suggests that we may restrict the Hilbert
space of the $N$ atoms to the subspace spanned by the ground state
$|g\rangle=|g_1,\dots,g_N\rangle$ and the single-excited-atom
states $|j\rangle=|g_1,\dots,e_j,\dots,g_N\rangle$ with
$j=1,\dots,N$. Hence, we set
\begin{equation}\label{psi}
    |\Psi(t)\rangle=\alpha(t)|g\rangle+e^{-i\Delta_0
    t}\sum_{j=1}^N\beta_j(t)|j\rangle
\end{equation}
where we will approximate $\alpha\approx 1$ after having evaluated
the different expectation values, e.g.
$\langle\hat\sigma_j\rangle\approx\beta_j$ and
$\langle\hat\sigma^\dagger_j\hat\sigma_m\rangle\approx
\beta^*_j\beta_m $. So, Eq.(\ref{s1:Mark}) yields
\begin{eqnarray}
   \frac{d\beta_{j}(t)}{dt} &=&
  \left(i\Delta_0-\frac{\Gamma}{2}\right)\beta_j(t)-\frac{i\Omega_0}{2}e^{i\mathbf{k}_0\cdot\mathbf{r}_j}-\frac{\Gamma}{2}
  \sum_{j\neq m}\gamma_{jm}
  \beta_m(t)\label{s1:beta},
\end{eqnarray}
with initial conditions $\beta_j(0)=0$. The self-interaction term,
$\Gamma\gamma_{jj}=\Gamma-i\Delta\Omega_{LS}$ yields the
single-atom spontaneous decay $\Gamma$ and the single-atom Lamb
shift $\Delta\Omega_{LS}$, which can be reabsorbed in the
definition of the atomic frequency $\omega_a$, and will be
neglected in the present analysis.

Considering the force applied to the center-of mass of the atomic
ensemble, $\hat{\mathbf{F}}=(1/N)\sum_j \hat{\mathbf{F}}_j$, from
Eqs.(\ref{Force-abs}) and (\ref{Force-emi:5}) the components along
the $z$ axis are
\begin{eqnarray}
    \langle\hat{F}_{az}\rangle&=& \hbar
    k_0\frac{\Omega_0}{N}\sum_{j=1}^N \textrm{Im}
    \left(
    e^{i\mathbf{k}_0\cdot \mathbf{r}_j}\beta_{j}^*
    \right)\label{Force-abs-beta}\\
    \langle\hat{F}_{ez}\rangle&=& -\frac{\hbar k_0\Gamma }{2N}\sum_{j\neq m}
    \hat{z}_{jm}
    j_1(k_0 r_{jm})i(\beta^*_j\beta_m-\textrm{c.c.}),
    \label{Force-emi-beta}
\end{eqnarray}
where $j_1(z)=\sin(z)/z^2-\cos(z)/z$ is the first order spherical
Bessel function and $\hat{z}_{jm}=(z_j-z_m)/r_{jm}$. Note also
that the self-force (\ref{self}) has zero average since
$\sum_{_{\mathbf{k}}} \mathbf{k}=0$ (although in general its
fluctuations are different from zero).

Also, from Eq.(\ref{E4}) it is possible to obtain the average
intensity of the scattered radiation as a function of the atomic
wave function,
\begin{equation}\label{Isca}
    I(\mathbf{r},t)=\epsilon_0 c\langle\hat E^\dagger(\mathbf{r},t)\hat E(\mathbf{r},t)\rangle=
     \left(\frac{d^2\omega_0^4}{16\pi^2\epsilon_0 c^3 r^2}\right)\left|\sum_{j=1}^N
    e^{-i\mathbf{k_{s}}\cdot \mathbf{r}_j}
    \beta_j(t)\right|^2.
\end{equation}
The state (\ref{psi}) may be conveniently expressed in the timed
Dicke (TD) basis, introduced originally by Dicke \cite{Dicke54}
and successively considered by R. Friedberg and coworkers
\cite{Friedberg} for their study on cooperative Lamb shift and by
M.O. Scully and coworkers \cite{Scully06,Scully09} to describe
cooperative decay of $N$ atoms prepared in a symmetric phased
state. The completely symmetric TD state is
$|+\rangle_{\mathbf{k}_0}=(1/\sqrt{N})\sum_j\exp(i\mathbf{k}_0\cdot
\mathbf{r}_j)|j\rangle$ and Eq.(\ref{psi}) can be written as
\begin{equation}\label{psi:TD}
    |\Psi(t)\rangle=\alpha(t)|g\rangle+
    e^{-i\Delta_0 t} \beta_{TD}(t)|+\rangle_{\mathbf{k}_0}+e^{-i\Delta_0
t}\sum_{s=1}^{N-1}\gamma_s(t)|s\rangle_{\mathbf{k}_0},
\end{equation}
where $|s\rangle_{\mathbf{k}_0}$ groups all the states orthogonal
to $|+\rangle_{\mathbf{k}_0}$ \cite{Scully06}.

A numerical analysis of Eq.(\ref{s1:beta}) shows that, for a constant driving field
$\Omega_0$ and for atomic cloud sizes much larger than the optical
wavelength, the occupation probability of the states
$|s\rangle_{\mathbf{k}_0}$ is only a small fraction of the atomic state \cite{Bux10}
and it is in general negligible, so that Eq.(\ref{s1:beta}) becomes
\begin{eqnarray}
   \frac{d\beta_{TD}}{dt} &=& -\frac{i}{2}\sqrt{N}\Omega_0
   +i\left(\Delta _0-\Delta_{N}\right)\beta_{TD}-
   \frac{1}{2}\Gamma Ns_N\beta_{TD}\label{s1:TD},
\end{eqnarray}
where
\begin{eqnarray}
    s_N &=& \frac{1}{N^2}\sum_{j,m=1}^N\frac{\sin(k_0|\mathbf{r}_j-\mathbf{r}_m|)}{k_0|\mathbf{r}_j-\mathbf{r}_m|}
    e^{-i\mathbf{k}_0\cdot(\mathbf{r}_j-\mathbf{r}_m)}=
    \frac{1}{4\pi}\int_0^{2\pi}d\phi\int_0^{\pi}d\theta\sin\theta\left|S_N(k_0,\theta,\phi)\right|^2\label{EqsN}\\
    \Delta_{N}&=-&\frac{\Gamma}{2 N}\sum_{j\neq
    m}^N\frac{\cos(k_0|\mathbf{r}_j-\mathbf{r}_m|)}{k_0|\mathbf{r}_j-\mathbf{r}_m|}e^{-i\mathbf{k}_0\cdot(\mathbf{r}_j-\mathbf{r}_m)}
    =-\frac{\Gamma N}{8\pi^2}\textrm{P}\int_0^{\infty}\frac{d\kappa\kappa^3}{\kappa-1}
    \int_0^{2\pi}d\phi\int_0^{\pi}d\theta\sin\theta\left|S_N(k_0\kappa,\theta,\phi)\right|^2\label{EqDN}
\end{eqnarray}
where $\kappa=k/k_0$,
\begin{equation}\label{EqStructurefactor}
    S_N(\mathbf{k})\equiv\frac{1}{N}\sum_{j=1}^Ne^{-i(\mathbf{k}-\mathbf{k}_0)\cdot\mathbf{r}_j}
\end{equation}
is the factor form and the integral over $\kappa$ in
Eq.(\ref{EqDN}) is evaluated as a principal part.
The term
$\Delta_{N}$ is the collective Lamb frequency shift
\cite{Friedberg,Scully09LS}. At steady state we find
\begin{equation}\label{EqSteadybeta}
    \beta_{TD} =\frac{\Omega_0\sqrt{N}}{2(\Delta_0-\Delta_{N})+iN\Gamma
    s_N}~,
\end{equation}
and
\begin{equation}
    \langle\hat{F}_{z}\rangle= \langle\hat{F}_{az}\rangle+\langle\hat{F}_{ez}\rangle=
    \hbar k_0\Gamma\frac{\Omega_0^2}{4(\Delta_0-\Delta_{N})^2+N^2\Gamma^2s_N^2}N\left[
    s_N-f_N\right]
    \label{Force-TD}
\end{equation}
where
\begin{eqnarray}
    f_N &=&\frac{1}{N^2}\sum_{j\neq m}
    \hat{z}_{jm}
    j_1(k_0 r_{jm})\sin(k_0 z_{jm}).
    \label{EqfN}
\end{eqnarray}

The cooperative radiation force can be obtained from the standard
single-atom radiation pressure force $F_1 = \hbar
k_0\Gamma\Omega_0^2/(4\Delta_0^2+\Gamma^2)$ substituting the
natural linewidth by the collective linewidth, $\Gamma_N=\Gamma Ns_N$, and
multiplying it by $1-f_N/s_N$, where $f_N/s_N$ is the probability
to observe the photon emitted in the forward direction. Isolating
the term $j=m$,
\begin{equation}\label{sN}
|S_N(\mathbf{k})|^2=\frac{1}{N}+ \sum_{j\neq
m}e^{i(\mathbf{k}_0-\mathbf{k})\cdot(\mathbf{r}_j-\mathbf{r}_m)}
\approx \frac{1}{N}+\left|S_\infty(\mathbf{k})\right|^2
\end{equation}
where the factor form $S_\infty(\mathbf{k})$ is evaluated for a
continuous approximation with density distribution
$n(\mathbf{r})$,
\begin{equation}\label{Sinf}
    S_\infty(\mathbf{k})=\frac{1}{N}\int_V d\mathbf{r} n(\mathbf{r})
e^{i(\mathbf{k}_0-\mathbf{k})\cdot\mathbf{r}}.
\end{equation}
Then, $s_N\approx (1/N)+s_\infty$ and $f_N\approx f_\infty$ where,
\begin{equation}\label{sf}
    s_\infty=\frac{1}{4\pi}\int d\Omega_{\mathbf{k}}
    |S_\infty(\mathbf{k})|^2,\quad\,\quad
    f_\infty=\frac{1}{4\pi}\int d\Omega_{\mathbf{k}}\cos\theta
    |S_\infty(\mathbf{k})|^2
\end{equation}
and Eq.(\ref{Force-TD}) becomes
\begin{equation}\label{PF2}
    F_z=\frac{\hbar k_0\Gamma\Omega_0^2 }{4(\Delta-\Delta_N)^2+\Gamma_N^2}
    \left[1+\frac{N}{4\pi}\int d\Omega_{\mathbf{k}}(1-\cos\theta)
    |S_\infty(\mathbf{k})|^2
    \right].
\end{equation}
The factor form $S_\infty(\mathbf{k})$ and the integrated factors
$s_\infty$ and $f_\infty$ have been calculated in ref. \cite{Courteille10}
for a Gaussian density distribution with ellipsoidal profile,
$n(\mathbf{r})_0\exp[-(x^2+y^2)/2\sigma_r^2-z^2/2\sigma_z^2]$,
yielding
$S_\infty(k_0,\theta)=\exp\{-\sigma^2[\sin^2\theta+\eta^2(\cos\theta-1)^2]/2\}$,
where $\sigma=k_0\sigma_r$ and $\eta=\sigma_z/\sigma_r$ is the
aspect ratio. For spherical and large clouds ($\eta=1$ and
$\sigma\gg 1$), $s_\infty\approx 1/(4\sigma^2)$, $f_\infty\approx
s_\infty-1/(8\sigma^4)$ and
the collective Lamb shift is $\Delta_N\approx\Delta_\infty$ where
(see \cite{Friedberg10b} and Appendix B)
\begin{equation}\label{deltaLS}
    \Delta_{\infty}=-\frac{\Gamma N}{4\sqrt{\pi}\sigma^3},
\end{equation}\label{EqLambShift}
which is a redshift, proportional to the number of atoms in a cubic wavelength
\cite{Friedberg}, i.e. atomic density and not optical thickness
$b_0=3N/\sigma^2$.
    \begin{figure}[t]
        \centerline{{\includegraphics[height=6cm]{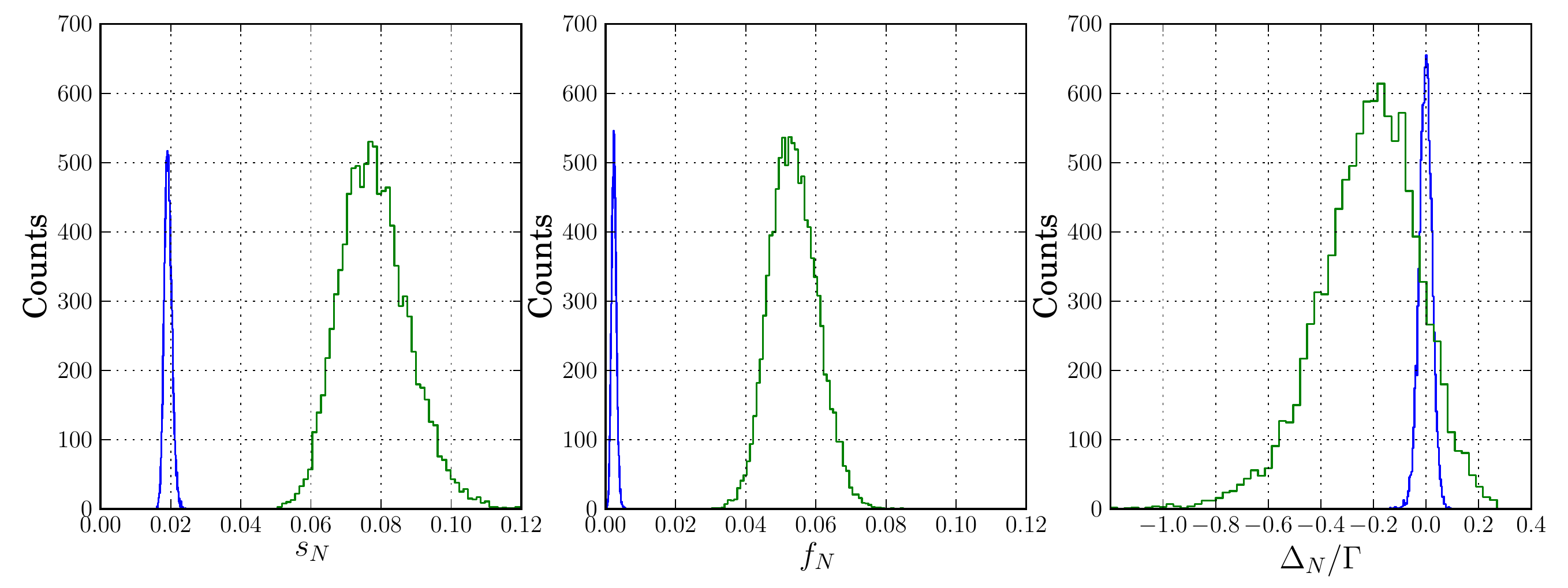}}}
        \caption{(color online) Distributions for values of $s_N, f_N$ and $\Delta_N$ for $N=50$ atoms, plotted for $10000$
        configurations for a size corresponding to $\sigma=10$ (blue curves) and $\sigma=2$ (green curves).}
        \label{FigFluctuations}
    \end{figure}
These values for $s_{\infty}, f_{\infty}$ and $\Delta_{\infty}$ can be compared to numerical evaluation of the $s_{N}, f_{N}$ and
$\Delta_{N}$ for a finite number of atoms and a specific configuration. In Fig. \ref{FigFluctuations} we show the distribution of
these values for different sample size.

\begin{figure}[t]
        \centerline{{\includegraphics[height=6cm]{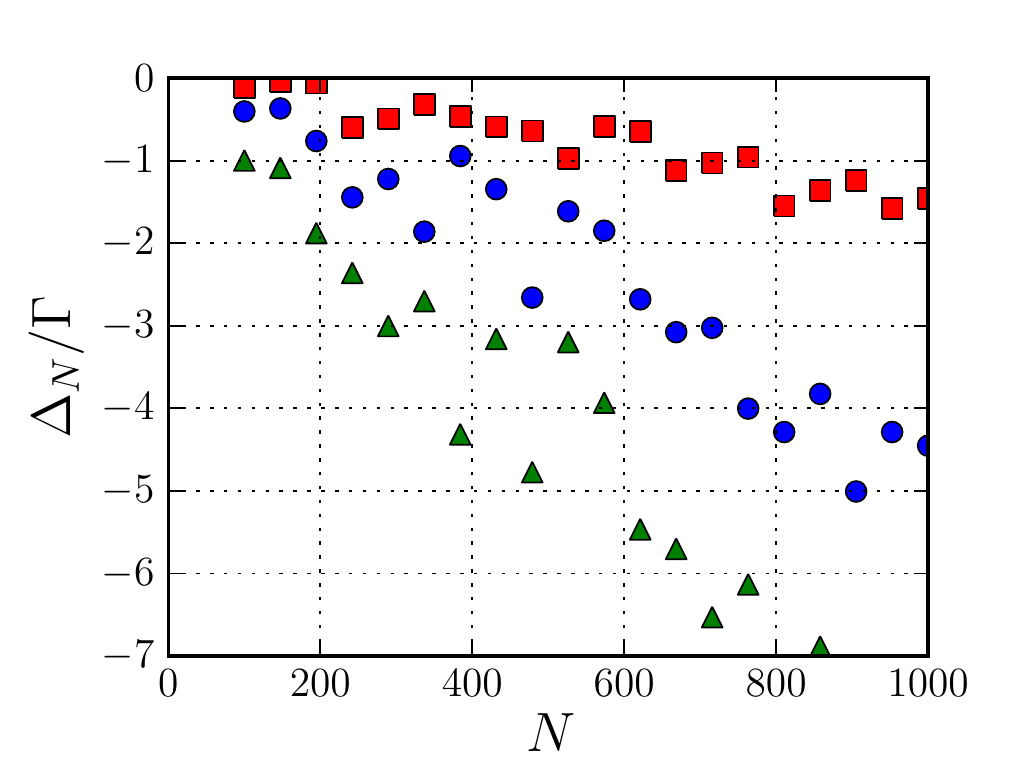}}}
        \caption{(color online) Collective Lamb shift vs atom number for $\sigma=1.6$ (green triangles)  $\sigma=2$ (blue circles) and $\sigma=3$
(red squares).}
        \label{FigCollectiveLamb}
    \end{figure}

In our numerical simulations shown in Fig. \ref{FigCollectiveLamb} we observe strong configuration dependent fluctuations for the value
of the collective Lamb shift. A precise comparison with our analytical expression, valid for large clouds, is thus cumbersome and did
not allow us to validate precise predictions of the numerical factor in Eq. (\ref{deltaLS}).

\begin{figure}[t]
        \centerline{{\includegraphics[height=6cm]{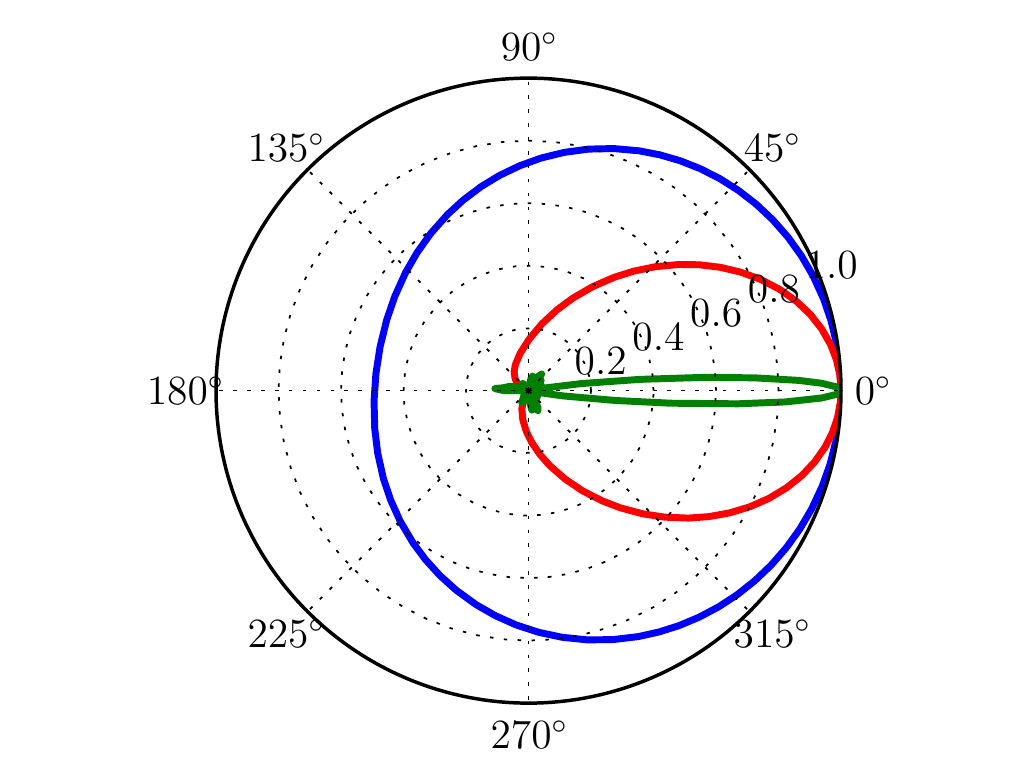}}}
        \caption{(color online) Emission diagram computed according to Eq. (\ref{Isca}) for the Timed Dicke state
        $|+\rangle_{\mathbf{k}_0}$ with $N=40$ atoms : $\sigma = 0.4$ (blue), $\sigma = 1$ (red) , $\sigma = 8$ (green).}
        \label{FigEmissiondiagram}
    \end{figure}

Normalizing the radiation pressure force with
respect to the single atom value, we obtain for large atomic
samples,
\begin{equation}
    \frac{\langle\hat{F}_{z}\rangle}{F_1}=
    \frac{4\Delta_0^2+\Gamma^2}{4(\Delta_0-\Delta_{N})^2+\Gamma^2(1+b_0/12)^2}\left[1+\frac{b_0}{24\sigma^2}\right]
    \label{FonF1}
\end{equation}
Finally, from Eq.(\ref{Isca}) we obtain the scattered intensity
\begin{equation}\label{Isca:TD}
    I(\mathbf{r})=
     \left(\frac{I_0}{16\pi^2 k_0^2 r^2}\right)
     \left[\frac{\Gamma^2}{4(\Delta_0-\Delta_{N})^2+\Gamma^2(1+b_0/12)^2}\right]
     \left[N+N^2|S_\infty(\mathbf{k}_s)|^2\right].
\end{equation}
This expression of the scattered intensity illustrates the role of the shape of the atomic cloud for the modified emission diagram.
The emission diagram of the TD state is shown in Fig. \ref{FigEmissiondiagram}. It illustrates the strong forward emission by the cloud when
its size exceeds a few optical wavelengths, reminiscent of Mie scattering, or more precisely of Rayleigh-Debye-Gans \cite{Hulst1957}.
As we will discuss in the following section, a modified emission diagram yields a modified radiation pressure force, as the recoil of the
scattered photon (partially) compensate the recoil effect at absorption.

\section{Heuristic approach}\label{HeuristicApproach}
The result (\ref{Force-TD}) can be
interpreted heuristically considering the momentum balance in a
given time interval $\delta t$ \cite{Dalibard}. During $\delta t$,
$N$ two-level atoms with positions $\mathbf{r}_j$ ($j=1,\dots,N$)
do $\delta N$ florescence cycles, each time absorbing a photon
with momentum $\hbar \mathbf{k}_0$ from the laser and emitting a
photon with momentum $\hbar \mathbf{k}_i$ ($i=1,\dots,\delta N$) in a random direction
$\mathbf{\Omega}_i$, with probability
$P_{i,j}=P(\mathbf{\Omega}_i,\mathbf{r}_j)$. The momentum
variation for the $j$th atom after $\delta N$ cycles is
\begin{equation}\label{dpa}
    \delta \mathbf{p}_j=(\hbar \mathbf{k}_0)\delta N -\sum_{i=1}^{\delta
    N}(\hbar \mathbf{k}_i)P_{i,j}.
\end{equation}
For a single isolated atom the emission is isotropic and
$P_{i,j}=1$, but for  $N$ atoms the emission can be not isotropic
depending on the atomic distribution. Also, the excitation could
be not uniform if the  phase front of the driving beam is getting
distorted by the  refractive index changes in the atomic cloud.
Assuming for simplicity that the excitation is uniform over the
entire atomic ensemble and neglecting phase distortion effects \cite{Gordon73, Ketterle05},
$\delta N$ will be the same for all the atoms and $|\mathbf{k}_i|=k_0$. Considering the
momentum variation along the direction of the incident photon ($z$
axis), after averaging over the atoms
\begin{equation}\label{dpCM}
    \delta p_{z}=\frac{1}{N}\sum_{j=1}^N
    \delta p_{j,z}=(\hbar k_0)\delta N - (\hbar k_0) \sum_{i=1}^{\delta
    N}P_{i} \cos\theta_i
\end{equation}
where $P_{i}=(1/N)\sum_{j}P_{i,j}=P(\cos\theta_i)$ is the emission
probability along the angle $\theta_i$. Considering $\cos\theta_i$
and $\delta N$ as independent random variables, the statistical
average of Eq.(\ref{dpCM}) is
\begin{equation}\label{dpave}
    \overline{\delta p_{z}}=(\hbar k_0)\,\overline{\delta N} - (\hbar
k_0)\, \overline{\delta N}
    \cdot\overline{ \cos\theta}
\end{equation}
where we assumed $\overline{\sum_{i}\cos\theta_i}\approx
\overline{\delta N}\cdot \overline{\cos\theta}$. Hence, the
pressure force is
\begin{equation}\label{PF}
    F_z=\frac{\overline{\delta p_z}}{\delta t}=(\hbar k_0)\left(\frac{\overline{\delta N}}{\delta
    t}\right)
    \left[1-\overline{\cos\theta}\right]
\end{equation}
Comparing with Eq.(\ref{Force-TD}) we found the following
correspondence
\begin{equation}\label{prob}
\left(\frac{\overline{\delta N}}{\delta
    t}\right)=\frac{\Omega_0^2 \Gamma_N}{4(\Delta-\Delta_{N})^2+\Gamma_N^2}
    \quad, \quad\quad
    \overline{\cos\theta}=\frac{f_N}{s_N}
\end{equation}
where $\Gamma_N=\Gamma N s_N$. So, the scattering rate
$(\overline{\delta N}/\delta t)$ is equal to the excitation
probability,
$\rho_{ee}=\Omega_0^2/[4(\Delta-\Delta_{N})^2+\Gamma_N^2]$, times
the collective decay rate, $\Gamma_N$. The radiation pressure
force (\ref{Force-TD}) is equal to the momentum photon, $\hbar
k_0$, multiplied by the scattering rate and by a geometrical factor
$1-\overline{\cos\theta}$ taking into account the directionality
of the scattered light. Cooperativity modifies both the scattering
rate, enhancing the decay rate and shifting the resonance
frequency, and the scattering direction. Small samples tend to
radiate isotropically whereas extended samples radiate
superradiantly in forward direction \cite{Courteille10,Prasad10}. These
cooperative effects can be revealed measuring radiation pressure
force by monitoring center-of-mass motion of large atomic clouds
released by magneto-optical traps \cite{Bienaime10,Bender10}, and then
identifying fast decay, shifts and modified emission diagrams
described by Eqs.(\ref{Force-TD}) and (\ref{Isca:TD}).

\section{Product state}\label{ProductState}
It has been noted that the same results
obtained for a symmetric TD state could be obtained assuming a
product state for $N$ atoms \cite{Eberly06,Friedberg10} (named
also
 `Bloch state' by some authors \cite{Friedberg07}):
\begin{equation}\label{state:product}
|\Psi(t)\rangle_{c}=\prod_{j=1}^N \left\{
\alpha_c(t)|g_j\rangle+\beta_c(t)e^{i \mathbf{k}_0\cdot
\mathbf{r}_j-i\Delta_0 t}|e_j\rangle \right\},
\end{equation}
where $\alpha_c(t)$ and $\beta_c(t)$ are the same for every atom,
with $|\alpha_c(t)|^2+|\beta_c(t)|^2=1$. The ansatz of
Eq. (\ref{state:product}) assumes each $j$th atom driven into the
excited state with equal probability $|\beta_c(t)|^2$ and phase
$\phi_j=\mathbf{k}_0\cdot \mathbf{r}_j-\Delta_0 t$. As it happens
for the symmetric TD state (\ref{psi:TD}), the driving field imposes
a coherence in the photons emitted spontaneously by each atom, so
that superradiance arises because the state is symmetric under
exchange of particles \cite{Scully:LP}. However, it is expected
that the quantum statistic of the symmetric TD  state will be
quite different from that of the 'quasi-classical' product state.
Notice that for $|\beta_c|\ll 1$ the product state
(\ref{state:product}) can be written in the following form
\cite{Friedberg07, Friedberg10b}
\begin{eqnarray}\label{state:appr}
|\Psi\rangle_{c}&=&\alpha_c^N|g\rangle+\alpha_c^{N-1}\beta_c\sum_{j}e^{i
\mathbf{k}_0\cdot \mathbf{r}_j-i\Delta_0 t} |j\rangle
+\alpha_c^{N-2}\beta_c^2\sum_{j\neq m} e^{i
\mathbf{k}_0\cdot(\mathbf{r}_j+\mathbf{r}_m)-2i\Delta_0 t}
|j,m\rangle+\dots
\end{eqnarray}
where $|j,m\rangle=|g_1,\dots,e_j,\dots,e_m,\dots,g_N\rangle$. Hence, the product state can be expanded in the
\textit{symmetric} TD states with $1$ to $N$ excited atoms. Only
in the limits $\alpha_c\approx 1$ and $|\beta_c|\ll 1$ the product
state reduces to the symmetric single-excited atom state
$|\psi\rangle\approx
|g\rangle+\beta_{c}\sqrt{N}|+\rangle_{\mathbf{k}_0}$ if only the
first two terms of Eq. (\ref{state:appr}) are retained. The
expectation values for the state (\ref{state:product}) are
$\langle\hat\sigma_j\rangle=\alpha_c^*\beta_c$ and
$\langle\hat\sigma_m^\dagger\hat\sigma_j\rangle=|\alpha_c|^2|\beta_c|^2$,
so for $\alpha_c\approx 1$ they coincide with those obtained from
the symmetric TD state. Differences between the product and the
symmetric TD states should appear when higher-order expectation
values are observed, as for instance
$\langle\hat\sigma_j\hat\sigma_m\rangle$, which is zero for the TD
state and $\alpha^{*2}\beta_j\beta_m\approx\beta_j\beta_m$ for the
product state. Notice that operator ordering produces different
results in high-order expectation values if scattered photons or
atomic forces are measured. These features  and  non classical
effects studies in cooperative scattering by cold atoms will be
the object of a future investigation.

\section{Conclusion}\label{Conclusion}
In this paper, we have included a more precise kernel to evaluate the cooperative radiation pressure force on a cloud of two-level systems.
The collective Lamb shift leads to a shift $\Delta_N$ of the resonance, which is proportional to the spatial density. As we have used a
scalar model in this paper, near field and polarization effects are neglected. One thus needs to consider this shift with some scepticism as
the numerical factor for this shift in a real system will be
strongly modified by the vectorial nature of the light \cite{Friedberg}. For dilute clouds, we recover previous results \cite{Bienaime10},
where these density effects are negligible. We also presented a simple model to estimate the radiation pressure force from the modified emission
diagram and assuming coupling to the single photon superradiant (Timed Dicke) state \cite{Scully06}. This approach can be useful to estimate not
only average forces but also fluctuations and dissipation. Finally, we noted that in the low intensity limit, the average result we derived for
the cooperative radiation pressure force can be obtained either by assuming a driven Timed Dicke state or a product state
\cite{Eberly06, Friedberg07,Friedberg10}, with no entanglement required. Looking for non classical features in cooperative scattering of light
by a cloud of two-level system thus requires studies of higher orders either by using higher intensities or looking at correlations or
fluctuations of the force.

\section{Acknowledgements}

We acknowledge fruitful discussions with E. Akkermans, P. Courteille, M. Havey, I. Sokolov and stimulating presentations on this topic at the
PQE 2011 conference.

\appendix

\section{Evaluation of the integral kernel in Eq.(\ref{s1ter})}\label{AppendixA}

Let's consider the last term in Eq.(\ref{s1ter}) and pass to the
continuous frequency approximation:
\begin{eqnarray}
  I(\mathbf{r}_{jm})=\sum_{\mathbf{k}}g_k^2
  e^{i\mathbf{k}\cdot\mathbf{r}_{jm}}
  \int_0^t dt'
  \hat\sigma_m(t-t')\,e^{-ic(k-k_0)t'}
  \rightarrow\frac{V_{ph}}{(2\pi)^3}
  \int d\mathbf{k}g_k^2
  e^{i\mathbf{k}\cdot\mathbf{r}_{jm}}
  \int_0^t dt'
  \hat\sigma_m(t-t')\,e^{-ic(k-k_0)t'}
  \label{K:1}.
\end{eqnarray}

We exchange the integration order and introduce spherical coordinates, $d\mathbf{k}=dk k^2 d\phi\,d\theta\sin\theta$.
After integration of the angular part, we obtain
\begin{eqnarray}
  I(\mathbf{r}_{jm})=\frac{V_{ph}}{2\pi^2}
  \int_0^t dt'
  \hat\sigma_m(t-t')e^{ick_0t'}
  \int_0^\infty dk k^2g_k^2
  \frac{\sin(k r_{jm})}{k r_{jm}}
  e^{-ick t'}
  \label{K:2}.
\end{eqnarray}
where $r_{jm}=|\mathbf{r}_{jm}|$. We approximate the $k$ integral
as
\begin{eqnarray}
  \int_0^\infty dk k^2g_k^2
  \frac{\sin(k r_{jm})}{k r_{jm}}
  e^{-ick t'}\approx \frac{k_0^2g_{k_0}^2}{2ik_0
  r_{jm}}\int_{-\infty}^\infty dk
  \left\{
  e^{-ick(t'-r_{jm}/c)}-e^{-ick(t'+r_{jm}/c)}
  \right\}
  \label{K:3},
\end{eqnarray}
where we made the following approximations: a) we assumed the
spectrum centered around $k\approx k_0$, so that $kg_{k}^2\approx
k_0g_{k_0}^2$; b) we extended the lower integration value from $0$
to $-\infty$, since the relevant values of $k$ are around $k_0$.
Using the expression above, we write
\begin{eqnarray}
  I(\mathbf{r}_{jm})=\frac{\Gamma}{2ik_0
  r_{jm}}
  \int_0^t dt'
  \hat\sigma_m(t-t')e^{ick_0t'}
  \left\{
  \delta(t'-r_{jm}/c)-\delta(t'+r_{jm}/c)
  \right\}=\frac{\Gamma}{2}\,
  \frac{e^{ik_0r_{jm}}}{ik_0
  r_{jm}}\hat\sigma_m(t-r_{jm}/c)
  \label{K:4}.
\end{eqnarray}
where $\Gamma=V_{ph}k_0^2g_{k_0}^2/(\pi c)$.
We observe that this approach does not require to assume the Markov approximation before solving the time integral, as in the standard
approach \cite{Svi10}.
On the contrary, this approach allows to obtain
the retarded (or not local) kernel, which, when the `rapid transit
approximation' is assumed, i.e. $\hat\sigma_m(t-r_{jm}/c)\approx
\hat\sigma_m(t)$, reduces to the exponential kernel of Eq.(8).

\section{Collective Lamb shift for a Gaussian distribution}

Let consider Eq.(\ref{EqDN}) for a continuous distribution:
\begin{eqnarray}
    \Delta_{\infty}&=&-\frac{\Gamma N}{8\pi^2}\textrm{P}\int_0^{\infty}
     \frac{d\kappa\kappa^3}{\kappa-1}
    \int_0^{2\pi}d\phi\int_0^{\pi}d\theta\sin\theta\left|S_\infty(\kappa,\theta,\phi)\right|^2.\label{B:1}
\end{eqnarray}
A spherical Gaussian distribution,
$n(r)_0\exp(-r^2/2\sigma_R^2)$, yields
$S_\infty(\kappa,\theta,\phi)=\exp[-\sigma^2(\kappa^2+1-2\kappa\cos\theta)/2]$,
where $\sigma=k_0\sigma_R$. Inserting it in eq.(\ref{B:1}) we
obtain
\begin{eqnarray}
    \Delta_{\infty}&=&-\frac{\Gamma N}{4\pi}\textrm{P}\int_0^{\infty}
     \frac{d\kappa\kappa^3}{\kappa-1}e^{-\sigma^2(\kappa^2+1)}
    \int_0^{\pi}d\theta\sin\theta
    e^{2\sigma^2\kappa\cos\theta}\nonumber\\
    &=&-\frac{\Gamma N}{8\pi\sigma^2}\textrm{P}\int_0^{\infty}
     \frac{d\kappa\kappa^2}{\kappa-1}\left[
     e^{-\sigma^2(\kappa-1)^2}-e^{-\sigma^2(\kappa^2+1)^2}
     \right]\nonumber\\
     &=&-\frac{\Gamma N}{8\pi\sigma^2}\textrm{P}\int_0^{\infty}
     d\kappa\left(\kappa+1+\frac{1}{\kappa-1}\right)\left[
     e^{-\sigma^2(\kappa-1)^2}-e^{-\sigma^2(\kappa+1)^2}
     \right]\nonumber\\
     &=&-\frac{\Gamma N}{8\pi\sigma^2}\textrm{P}\int_{-1}^{\infty}
     dx\left(2+x+\frac{1}{x}\right)\left[
     e^{-\sigma^2x^2}-e^{-\sigma^2(2+x)^2}
     \right].
     \label{B:2}
\end{eqnarray}
For $\sigma\gg 1$ it is approximated by
\begin{eqnarray}
    \Delta_{\infty}
     &\approx &-\frac{\Gamma N}{8\pi\sigma^2}\textrm{P}\int_{-\infty}^{\infty}
     dx\left(2+\frac{1}{x}\right)
     e^{-\sigma^2x^2}\approx
     -\frac{\Gamma N}{4\sqrt{\pi}\sigma^3},
     \label{B:3}
\end{eqnarray}
in agreement with the result of Friedberg and Manassah
\cite{Friedberg10b}.


\begin{thebibliography}{10}
\bibitem{velocity} Light may propagate in dense atomic samples with a group velocity smaller than $c$ \cite{Labeyrie03}. In these case
the Markov approximation should be satisfied by a more stringent
condition.
\end{thebibliography}
\end{document}